# Activation energy spectrum for relaxation and polyamorphism in an ultra-viscous metallic glass former

Isabella Gallino[1,*], Daniele Cangialosi[2], Zach Evenson[3], Lisa Schmitt[1,4], Simon Hechler[1,5], Moritz Stolpe[1], and Beatrice Ruta[5,6]

[1] Chair of Metallic Materials, Saarland University, Campus C6.3, 66123 Saarbrücken, Germany

[2] Materials Physics Center (CFM/MPC), Paseo Lardizabal 5, 20018 San Sebastian, Spain

[3] Heinz Maier-Leibnitz Zentrum (MLZ) and Physik Department, Technische Universität München, Lichtenbergstrasse 1, 85748 Garching, Germany

[4] fem Research Institute for Precious Metals & Metals Chemistry, Katharinenstrasse 17, 73525 Schwäbisch Gmünd, Germany

[5] ESRF—The European Synchrotron, CS40220, 38043 Grenoble, France

[6] Institute of Light and Matter, UMR5306 Université Lyon 1-CNRS, Université de Lyon, 69622 Villeurbanne Cedex, France.

*corresponding author: Isabella Gallino. E-mail address: i.gallino@mx.uni-saarland.de. Tel: +49 (0)681 302 2052

**Abstract:** Many glass-formers exhibit phase transitions between two distinct liquid states. For some metallic glass-formers, the liquid-liquid transition is experimentally found in the supercooled liquid at intermediate temperature between the melting point and the glass transition temperature $T_g$. We report here on a liquid-liquid transition in an ultra-viscous metallic glass-former, accessed during long-time annealing. This study is conducted on the $Au_{49}Cu_{26.9}Si_{16.3}Ag_{5.5}Pd_{2.3}$ composition with a liquid-liquid transition temperature slightly lower than $T_g$. The consequence is that the high-temperature kinetically *fragile* liquid freezes into the glass during conventional processing and the underlying liquid-liquid transition is thus accessed by the system during annealing below $T_g$. Upon reheating, the reverse transformation is observed by calorimetry. This conclusion is supported by a broad collection of complementary laboratory and synchrotron-based techniques, such as differential- and fast- scanning calorimetry, and x-ray photon correlation spectroscopy. Our findings support the 'big-picture' proposed by Angell that liquids with different fragility occupy different flanks of an underlying order-disorder transition. Furthermore, our multiscale analysis reveals the existence of multiple decays of the enthalpy recovery, which is reflected in the observed microscopic ordering and aging mechanism of the glass through distinct stationary regimes interconnected by abrupt dynamical aging regimes.

*Keywords:* bulk metallic glass, aging, X-ray photon correlation spectroscopy, enthalpy recovery, liquid-liquid transition

## 1. Introduction

Several glass-formers present polyamorphic phase transformations between two liquids with different entropy and structure. These liquid-liquid transitions (LLT) are analogous to polymorphic transformations in crystalline materials, and encompass many glass-forming systems, i.e. multi-component alloys [1-5], water and aqueous solutions, $Y_2O_3$-$Al_2O_3$, $SiO_2$, $BeF_2$, thriphenyl phosphite, Si, P, and Ge among others [6-13]. In the 'big picture' proposed by Angell in his Turnbull Lecture [14], the two liquids involved occupy different flanks of an underlying order-disorder transition and have distinct kinetic fragility. The kinetic fragility can be taken from the empirical Vogel-Fulcher-Tammann (VFT) equation,

$$f(T) = f_0 exp\left(\frac{T_0 D^*}{T-T_0}\right) \qquad [1],$$

where $f$ is either viscosity η or relaxation time τ. $T_0$ is the putative temperature, at which the barrier to viscous flow would become infinite. The combined parameters $T_0$ and $D^*$ thus effectively model the temperature dependence of the liquid relaxation kinetics; i.e. the kinetic fragility. The larger the $D^*$, the 'stronger' the liquid, which more closely obeys the Arrhenius law in the limit $T_0 \rightarrow 0$. Kinetically fragile liquids, on the other hand, are characterized by smaller values of $D^*$.

The numerous aforementioned studies of liquid-liquid transitions have shown that, at high temperatures, kinetically fragile and less structurally ordered liquids can, upon cooling, transform spontaneously into a more ordered, kinetically stronger liquid without changing chemical composition, thereby leading to a fragile-to-strong polyamorphic transformation. According to Bragg-Williams theory, the higher the ordering energy, the higher the entropy change that accompanies an order-disorder transition, and the higher the transition temperature [15-16].

Classical kinetically strong network liquids like $SiO_2$ and $BeF_2$ may undergo such order-disorder transitions at very high temperature far above the melting point at a transition temperature $T_{LLT}$ ~ $3T_g$ [3, 14, 17-19], where $T_g$ is the glass transition temperature observed on the typical laboratory timescale. In the case of highly fragile liquids, the $T_{LLT}$ is located at such a low temperature that the system may undergo structural arrest before the transition [14]. Very recently, it was also experimentally observed that, in the case of thin films of a fragile polymer, the system can eventually fall into the energy minimum corresponding to that of the ideal glass [20], before the fragile-to strong transition can occur. For water, Si, Ge and some bulk metallic glass (BMG) formers, the fragile-to-strong transition is found at intermediate temperature between $T_g$ and the melting point, $T_m$ [3-4, 12-14]. In the case of BMG-forming liquids, many display kinetically fragile liquid behavior above $T_m$ and stronger liquid behavior when equilibrated below $T_g$ [2, 21-22]. This is usually taken as indirect evidence indicating that the system undergoes structural arrest before the LLT, and that the fragile-to-strong transition may indeed occur in the ultra-viscous state (where η > $10^{12}$ Pa s and τ > 100 s), which is usually experimentally inaccessible due to the long timescales associated to the dynamics and eventual intervening crystallization processes. In fact, indications of such transformations in the ultra-viscous state have been recently reported as anomalous behaviors in sub-$T_g$ enthalpy relaxation studies in Au-Ag-Pd-Cu-Si, Cu-Zr-Al, and La-Ce –based metallic glasses [23-25].

In this work, by combining microscopic and macroscopic experimental techniques, we find direct evidence of a fragile-to-strong transition below $T_g$ by studying the sub-$T_g$ relaxation behavior of $Au_{49}Ag_{5.5}Pd_{2.3}Cu_{26.9}Si_{16.3}$. This composition is robust against crystallization on long timescales [26] and the $T_m$ is low enough that the glass transition is observed upon cooling from the stable melt by means of fast scanning calorimetry (FSC). Our study also presents a multiscale investigation of the activation energy spectrum for relaxation upon applying a wide range of annealing times, from less than one second up to several days, during annealing at temperatures 50 to 20 K below $T_g$, where it is expected to encounter weakly coupled frozen-in relaxation processes [27]. Thermodynamic and kinetic analyses are performed using FSC [28], conventional differential scanning calorimetry (DSC), and thermal mechanical analyses (TMA) [29].

The results from these laboratory experiments are compared with detailed microscopic information on the collective atomic motion obtained using X-ray Photon Correlation Spectroscopy (XPCS) [30]. Our results reveal distinct multiple decays in the enthalpy recovery behavior, which correspond to states of local and transient equilibrium with increasingly higher activation energies. These results are confirmed at the atomic level, where we indeed observe different regimes of stationary dynamics separated by intermittent temporary aging regimes. In the framework of the potential energy landscape approach [31], we ascribe the observed relaxation process to transitions from a high-energy local minimum to energetically lower, more deeply relaxed states.

## 2. Results

### a. Enthalpy recovery

During annealing, the glass drifts towards a state of lower free energy and the system releases enthalpy. The released enthalpy is recovered back into the system when the relaxed glass is heated up through the glass transition, manifesting itself as an endothermic overshooting of the heat flow signal. Figure 1 shows the time evolution of the enthalpy recovery of $Au_{49}Ag_{5.5}Pd_{2.3}Cu_{26.9}Si_{16.3}$ after annealing below $T_g$. The ΔH(rec,t) and the ΔH(tot) values are obtained from the integration of the endothermic overshooting signal with respect to the heat flow signal of the standard treated material (see materials and methods for further details). Examples of up-scans of glasses relaxed for different annealing times $t_a$ at 373 K are shown in Fig. 1a. For values of $t_a$ up to 10,000 s, the onset of the endothermic overshoot is the same as the $T_g$ of the reference state, which is the standard treated glass. After 10,000 s, the onset shifts to higher temperatures, reflecting the transition toward a state with higher kinetic stability of the relaxed glass on a long timescale.

The equilibrium recovery behavior can be described by stretched exponential [Kohlrausch-Williams-Watts (KWW)] functions. This approach neglects non-linearity effects, that is, the structure dependence of the relaxation time [32-36]. However, the treatment based on a single KWW function is not sufficiently adequate for the treatment of the data of Fig. 1b. In fact, we do observe a separation of enthalpy recovery processes as the annealing temperature decreases. Only the high temperature data at 378 K can be adequately reproduced with a single stretched-exponential decay, with stretching parameter $β < 1$, i.e.

$$f(t) = \exp(-(t/\tau)^\beta) \qquad [2].$$

At lower temperatures the decay clearly splits into different steps, which we model as a sum of simple exponential decays:

$$f(t) = A + \sum_{i=1}^{N} h_i \exp(-t/\tau_i) \qquad [3].$$

Here, A is the baseline. The values of $h_i$ and $\tau_i$ correspond to the relaxation amplitude and to the characteristic relaxation time of the observed enthalpy decay, respectively. We wish to note here that each data point shown in Fig. 1b corresponds to a single measurement sample. The fitting procedure assured that always the same identical relaxation decays were identified independently of the number of the exponential terms assumed. Figure 1b shows the best fits as continuous lines obtained with equation 3, while the corresponding parameters are reported in the SI, together with the value for the activation energy $Q_i$ of each decay. The latter are obtained by fitting the temperature dependence of the resulting $\tau_i$ to the Arrhenius equation. The values of $\tau_i$, named hereafter as $\tau_i(\Delta H_{rec})$ and $Q_i$ are given in the inset of Fig. 1b.

**b. Isothermal enthalpy released**

The kinetics of the enthalpy released during relaxation of the standard treated $Au_{49}Ag_{5.5}Pd_{2.3}Cu_{26.9}Si_{16.3}$ glass was directly studied by analyzing the calorimetric signal recorded during isothermal DSC measurements between 368 and 390 K. The isothermal heat flow signals are integrated as a function of time to obtain the ΔH(released,t) functions as in [37]. The analysis is reported in the SI and the resulting equilibrium characteristic relaxation times $\tau(\Delta H_{released})$ are plotted in Fig. 5 as open red squares and discussed below.

**c. Isothermal viscosity change**

Experimentally determined viscosities for the as-cast $Au_{49}Ag_{5.5}Pd_{2.3}Cu_{26.9}Si_{16.3}$ glass in the vicinity of $T_g$ were measured by isothermal three-point beam-bending, and are reported here in the SI. The values of the equilibrium liquid viscosity $\eta_{eq}$ at each annealing temperature were obtained using the procedure described in Refs. [29, 22]. These values are plotted as a function of annealing temperature in Fig. 5 where they are fitted to the VFT equation (Eq. 1).

**d. Specific heat capacity**

Figure 2 shows the specific heat capacity, $C_p(T)$, of the glassy (open squares), crystalline (open triangles) and equilibrium supercooled liquid (SCL) (open circles) states of the $Au_{49}Ag_{5.5}Pd_{2.3}Cu_{26.9}Si_{16.3}$. The data are measured by applying an isothermal step-method as described in the methods. In addition, individual isothermal undercooling experiments were performed to measure the $C_p(T)$ of the liquid below the liquidus at temperatures between 565 and 630 K. The continuous curves correspond to the temperature dependence of $C_p$ for the equilibrium liquid, the crystalline state and the glassy state as determined through fitting of the experimental data to the empirical equations [38-39]:

$$C_p(\text{liquid}, T) = 3R + aT + bT^{-2} \qquad [4]$$

$$C_p(\text{crystal}, T) = 3R + cT + dT^2 \qquad [5]$$

$$C_p(\text{glass}, T) = \frac{3R}{M}\left(1 - \exp\left(1.5\frac{T}{T_D}\right)\right) \qquad [6]$$

where R is the ideal gas constant and a-d are fitting constants, $T_D$ is the Debye temperature and M a scaling factor close to unity [39]. The fitting parameters are listed in the SI.

Figure 2 also shows the values of $C_p(T)$ of the deeply undercooled liquid in its ultra-viscous state, as obtained from the enthalpy recovery values measured after long-time annealing below $T_g$ (shaded circles). The determination of these values is described in the methods, and follows the methodology applied in [40, 37]. These values for the equilibrated SCL on a long timescale represent two distinct, albeit metastable, thermodynamic equilibrium states. The $\Delta C_p$(SCL-crystal, $T_g$) values for these two equilibrium states are $14.6 \pm 2.0$ J g-atom$^{-1}$ K$^{-1}$ and $20.3 \pm 2.0$ J g-atom$^{-1}$ K$^{-1}$ for the blue and the grey symbols, respectively.

### e. Atomic dynamics during annealing below $T_g$

The occurrence of intermittent aging regimes is revealed here by microscopic studies of the atomic dynamics in the glassy state using XPCS during long-time isothermal anneals below $T_g$. We observe this to go hand-in-hand with the existence of different decays in the time evolution of the enthalpy recovery. The XPCS technique provides information on the microscopic time necessary for structural rearrangements at the atomic level [30, 41]. Data taken with an $Au_{49}Cu_{26.9}Si_{16.3}Ag_{5.5}Pd_{2.3}$ as-spun ribbon during isothermal annealing at different temperatures from room temperature up to $T_g$ reveal that the collective atomic motion is persistent and stationary, with no immediate signs of physical aging. However, after long-time regimes of approximately $10^3$ to $10^4$ s of stationary dynamics, sudden physical aging events are observed to occur, reflected in an abrupt slowdown of the atomic dynamics.

Figure 3a shows an example of this behavior for the annealing step at 377 K by reporting the direct correlation between a given atomic configuration and its spontaneous temporal evolution, the so called two-times correlation function TTCF [42]. The center diagonal from the bottom left corner to the upper right corner corresponds to the elapsed time of the measurement. The width of the yellowish diagonal contour is proportional to the characteristic timescale on which any given atomic configuration no longer correlates to that measured at a later time and it thus directly relates to the structural relaxation time τ. If the width becomes larger with time, the corresponding atomic dynamics decelerates, which is a characteristic feature of physical aging. In Fig. 3a, the width of the diagonal contour is constant for about 5,500 s until a transitory aging regime occurs, as marked by the sudden broadening of the intensity profile of the TTCF at about $t_a$ = 6,040 s. The TTCF in Fig. 3b corresponds to the continuation of the measurement at T = 377 K shown in Fig. 3a.

Quantitative information on the microscopic dynamics can be obtained by extracting from the TTCF the standard intensity-intensity correlation functions, $g_2(q_0,t)$, for the two stationary regimes prior to and after the aging regime. These are shown in Fig. 3c together with the fit using a KWW expression of the form

$$g_2(q_0, t) = 1 + c \exp[2(t/\tau)^\beta] \qquad [7].$$

Here, τ = τ(XPCS) is the relaxation time measured in XPCS, β is the shape parameter and the pre-factor $c = \gamma f_q^2$ is the product between the experimental contrast γ and the square of the Debye-Waller factor $f_q$ [41].

The elapsed time of the aging regime is ~ 500 s, during which a steady slowdown of the atomic dynamics is observed, similar to Refs. [41-43]. In this case, thus, the continuous microscopic aging occurs through transitions between distinct stationary regimes. In fact, after the observed regime of physical aging, the atomic dynamics becomes overall slower – marked by an increase in τ from 570 ± 26 to 1650 ± 7 s –but remains persistently constant afterwards within the experimental timescale. The stationary regime labeled with b was observed to persist for at least 7,050 s.

Long-time stationary dynamics is observed at all other annealing temperatures of the XPCS experiment. Aging events, similar to that reported in Fig. 3, are observed only at temperatures below 377 K. Above this temperature, the atomic dynamics are stationary and persistent for annealing times many orders of magnitude greater than the corresponding XPCS relaxation times. For this high-temperature regime, the corresponding normalized $g_2(q_0,t)$ functions are shown in the plot of Fig. 4, while the resulting τ(XPCS,T) data are shown in the inset. The τ(XPCS,T) values are best fitted to an Arrhenius equation with an activation energy of 4.03 eV, thus comparable to that reported in Fig. 1b for the second-to-last step in enthalpy relaxation.

The transition from the glass to the supercooled liquid (SCL) is finally observed during continuous heating from 393 to 395 K, and it corresponds to a sudden increase of τ and to an abrupt change of the shape parameter of the $g_2(q,t)$ from β ≈ 1.8 towards values close to unity. This dynamical crossover has been also reported in other metallic glasses [41] and is further discussed in Ref. [44] together with the corresponding dynamics in the ultra-viscous SCL. The fact that the SCL is finally reached at this temperature is not a surprise considering that 396 K corresponds to the calorimetric $T_g$ in heating [45-46].

### 3. Discussion

**a. Activation energy spectrum of relaxation modes**

The atomic dynamics in the glassy state of $Au_{49}Cu_{26.9}Si_{16.3}Ag_{5.5}Pd_{2.3}$ displays temporally intermittent aging behavior similar to that observed recently in other fragile BMG systems based on Pd where the aging mechanism involves avalanche-like relaxations from a high energy local minimum to a deeper relaxed state [43, 47-48]. Differently from those cases, however, aging here only occurs on relatively short timescales and is accompanied by long stationary regimes. We interpret this behavior as the system's ability to visit several local and distinct metastable states during aging, reflecting a hierarchical structural relaxation process that does not involve appreciable rearrangement of an atomic backbone structure comprised mainly of the larger, slower moving atoms (e.g. Au or Pd). The observed persistent stationary atomic dynamics is indicative of a higher resistance to diffusive processes, and could be related to microscopic ordering processes not involving density changes [43].

The high-temperature glass configuration that is observed during aging at temperatures between 377 and 391 K (see Fig. 4), can be interpreted as a state trapped in a deep local energy minimum. Here, even though the atomic dynamics becomes increasingly faster as the temperature increases [τ(XPCS) ~ 10 s], we did not observe any sign of physical aging as a function of $t_a$ towards an even deeper glassy state or towards the equilibrium state. The thermal energy is apparently not enough to immediately overcome the high local activation energy barriers during a single isothermal treatment. Only upon approaching the calorimetric $T_g$ the backbone structure as a whole starts to flow, facilitating the transition back into the ultra-viscous SCL. This surprising behavior strongly contradicts macroscopic measurements where aging close to $T_g$ occurs on timescales similar to the intrinsic relaxation time, and strengthens the idea of a strong length scale dependence of the dynamics in the glassy state.

The activation energy spectrum of relaxation modes reported here for $Au_{49}Cu_{26.9}Si_{16.3}Ag_{5.5}Pd_{2.3}$ seems to be a universal feature of glass-forming systems, independent of the nature of the relaxation processes. Multiple relaxation behaviors were observed not only in metallic glass-forming systems (e.g. [24, 47-51]), but also in numerous studies on fragile polymeric glass formers [52-59]. Studies carried out on fragile As-Se glasses at low temperatures report a multi-step enthalpy recovery behavior [60], similar to the results reported in Fig. 1. From a theoretical viewpoint, a distribution in the logarithm of relaxation times, for example, arises within the random first order transition theory of the glass transition [61], with the separation of the peaks of the relaxation distribution becoming more pronounced at lower temperatures. This distribution differs from the mostly single peaked distributions found in equilibrium supercooled liquids [62-64], and cannot be explained in terms of β-relaxation-type of atomic rearrangements in the glassy state, which are confined mainly to short-range order length scale (less than ~ 5.5 Å) [45, 65-66].

**b. Atomic mobility in the ultra-viscous state**

The relaxation behavior of BMGs can be linked to atomic mobility, which can be directly probed, for example, in radiotracer experiments. A diffusion process in metallic glasses on the order of one atomic displacement can involve thermally activated, highly cooperative atomic motions of 10 to 20 atoms in a chainlike manner [67]. In addition, BMG-formers are typically multi-component systems characterized by large atomic mismatch among the atomic species. Thus, it is expected that, at low temperature where there are fewer active degrees of freedom, some of the more sluggish processes, controlled by the slower diffusive species, remain frozen, leading to a more visible separation of the relaxation modes. The presence of multiple decays in the enthalpy change during aging could be justified, therefore, by relaxation processes involving cooperative structural rearrangements that involve atoms of predominately small and intermediate sizes. These processes do not necessarily involve the motion of the entire glassy matrix, the cause for macroscopic plastic flow. On the other hand, these intermediate timescale relaxation processes are much slower than the β-relaxation process, which is attributed to fast, string-like motions of atoms [62-64].

Our finding of a distinct activation energy spectrum of enthalpy relaxation modes is in agreement with a few radiotracer experiments performed with Pd-based and Zr-based multicomponent glass forming alloys [68-72, 63]. The diffusivity of the large atoms like Pd, Au and Zr, is found to decouple from the diffusivity of the smaller components. Only for the large atomic species does the relation between the equilibrium viscosity and diffusion coefficient D, ($D \propto T/\eta$), hold over an entire 14 orders of magnitude change in diffusivity [70, 63]. The smaller components are seen to strongly deviate from the Stoke-Einstein relation below a certain temperature. This is the case for P, Si, and highly polarizable atoms like Cu, Co, and Cr. Thus, the apparent viscosity values extracted from the self-diffusion coefficients of these very mobile atoms would fall below the macroscopic equilibrium viscosity data, which is well represented by the diffusion of large-size components. The difference increases monotonically with decreasing temperature, and at $T_g$ the difference is some 4 orders of magnitude [70, 63]. Although radiotracer experiments have not been yet performed with the $Au_{49}Cu_{26.9}Si_{16.3}Ag_{5.5}Pd_{2.3}$ system, it is likely that similar tremendous asymmetries in the mobility of the different species during aging must be present in this particular system. We argue that the observed intermediate timescale relaxation processes are induced by the asymmetric diffusion of Cu enhanced by the presence of Si as is discussed elsewhere as well [73, 47].

**c. The fragile-to-strong crossover in the ultra-viscous state**

Apart from the robust evidence of the presence of multiple mechanisms in the enthalpy relaxation process, the main novelty of this work lies in the detection of the thermodynamic signature of a fragile-to-strong transition in the ultra-viscous state below the standard calorimetric $T_g$. We have also recently observed, at the atomic level, the LLT in the $Au_{49}Cu_{26.9}Si_{16.3}Ag_{5.5}Pd_{2.3}$ SCL upon equilibrium cooling [44]. Differently from that work, here the LLT is observed at the macroscopic level upon annealing from the glass. As a result, the SCL state obtained on a long timescale below a certain transition temperature is thermodynamically more stable than the state obtained above. Through annealing, we are able to experimentally access this state, which is evidenced in Fig. 2 by the different values of $C_p$ for the SCL, estimated from enthalpy recovery experiments after equilibration (blue vs. grey shaded circles). In terms of the jump of $C_p$ at $T_g$, the low-temperature equilibrated state (blue shaded circles) reflects a stronger thermodynamic fragility behavior than the high-temperature equilibrated state (grey shaded circles), in agreement with the thermodynamic fragility concept of Angell [74]. This difference in $C_p$ is expected also to reflect distinct underlying kinetic fragility values for each state [75-76], which we quantify in the following.

We quantify in Fig. 5 the kinetic fragility of the two distinct liquids discussed above using an Angell-type, or fragility, plot that represents the temperature dependence of the measured relaxation time. The data are here rescaled as a function of (389 K)/T, where 389 K is the $T_{LLT}$ associated with the fragile-to-strong crossover at the atomic level as detected in [44]. Here, the isothermal equilibrium viscosities (filled blue circles), the relaxation times obtained for the decays of the enthalpy release, $\tau(\Delta H_{released})$ (open red squares), and the final enthalpy recovery relaxation time all show a similar VFT temperature-dependence. The VFT fit shown in Fig. 5 with $D^* = 20.9 \pm 0.4$ (dashed line) was obtained considering only the $\eta_{eq.}$ data (blue filled circles), and is characteristic of a relatively strong BMG-forming liquid [75]. The two $\tau(\Delta H_{released})$ data points (red open square) measured at 388 and 390 K follow a steeper trend, which is in agreement with the more fragile behavior of the DMA alpha relaxation time, $\tau_\alpha$ [44-45] (open triangles), and the $\tau$(XPCS) data in the SCL (open red circle symbols at 395 and 396 K). These three sets of data are fit together to the VFT equation. The obtained $D^* = 9.78 \pm 0.4$ corresponds to one of the most kinetically fragile BMG-forming liquids detected in the vicinity of $T_g$ [75]. The open crossed circles in Fig. 5 are taken from the calorimetric data of Ref. [25] and actually appear to be characteristic of two distinct systems, in agreement with the extrapolations of our VFT fits.

In Fig. 5, the Arrhenius fit obtained from the $\tau$(XPCS) in the glassy state agrees well with the activation energy of the second-to-last enthalpy recovery decay (see top dashed-dot line). The VFT and the Arrhenius curves shown in Fig. 5 are also reproduced in Fig. 6, which corresponds to a temperature-time-transformation (TTT) diagram for $Au_{49}Ag_{5.5}Pd_{2.3}Cu_{26.9}Si_{16.3}$. The temperature dependence of the two data points corresponding to $\tau_5$, obtained from the last enthalpy recovery decay (open diamonds symbols), appears in Fig. 5 to be comparable to the temperature dependence of the equilibrium viscosity data (filled blue circles). Even from a cursory inspection, it is clear that these two points would have a higher activation energy than all other diamond symbols, and that viscous flow might well be connected to those structural rearrangements responsible for the slowest relaxation mode detectable in our enthalpy recovery decay analysis.

As briefly mentioned, a microscopic signature of the LLT has been recently reported by us in ref. [44] by combining advanced structural and dynamical synchrotron measurements using the same identical step- protocol applied to the aforementioned DSC experiment. The resulting relaxation times measured during cooling from the fragile liquid are shown in Fig. 6 as blue open triangles. They show a distinct change in the VFT temperature-dependence at 389 K from a D* value of 8.9 to 23.7, which marks the transition from a kinetically highly fragile to a stronger liquid and is reflected by an anomalous temperature dependence of the first peak position of total structure factor function [44].

**d. Reversibility of the fragile-to-strong crossover**

We argue that the endothermic overshoot signal (see Fig. 1a) after final equilibration has two contributions. The first is due to the enthalpy recovery event of a highly relaxed structure; the second is the endothermic signature of the polyamorphic strong-to-fragile transition, i.e. the reverse transition of the LLT that occurred during annealing below $T_{LLT}$ [3, 11, 77]. If we consider the annealing at 373 K, the maximum amount of enthalpy recovered due solely to the relaxation of the fragile glass is $200 \pm 25$ J g-atom$^{-1}$, which is the value obtained after an annealing for $t_a$ =10,000 s. If we subtract this value from the total area of the endothermic overshoot after equilibration ($270 \pm 10$ J g-atom$^{-1}$), the difference is the enthalpy contribution of the reverse transition to the endothermic overshoot, $\Delta H(LLT) = 70 \pm 25$ J g-atom$^{-1}$. In terms of restored disorder, it corresponds to an entropy change $\Delta S(ordering) = 0.187 \pm 0.06$ J g-atom$^{-1}$ K$^{-1}$, if we take 373 K as a reference. This corresponds to 2.4 % of the entropy of fusion. In the inset of Fig. 2 we compare this value to the entropy change associated to the liquid-liquid transition of the Vit106a and the Vit1 BMG systems from [3-4]. The $\Delta S(ordering)$ is linearly proportional to the temperature of the crossover between the two liquid states.

The endothermic overshoot of the enthalpy recovery starts to shift to higher temperatures as soon as (and not before) the fragile-to-strong transition begins to occur in the ultra-viscous state. The maximum temperature shift is obtained when the entire volume of the specimen has transformed, which in this study is determined as 10 K. In ref. [25], a shift of the endothermic overshoot to higher temperatures is also observable and even more pronounced than in our case (~28 K) due to the applied heating rate of the up-scan of 1000 K s$^{-1}$ – some four orders of magnitude higher than our DSC experiments. In fact, according to the extrapolation of the VFT-fits in the TTT diagram of Fig. 6, upon re-heating, the onset of the glass transition of the strong system occurs at a higher temperature than that of the fragile system, and the onset temperature difference increases as $\tau$ decreases. For this reason, the maximum shift of the onset of the endothermic overshoot is greater when a faster scanning rate is applied.

During the relaxation experiments carried out in this work, the long-time annealing seems to result in an equilibrium state that has the same thermodynamic stability as a liquid cooled slowly enough to maintain an internal equilibrium, in agreement with [77]. The fact that the two liquid states have reached the same level of thermodynamic stability is supported by calorimetry, as the area under the endothermic signals detected from both the specimen after the step- thermal protocol (dashed curve in Fig. 1a), and from the fully equilibrated glass at 373 K corresponds to an equal value of enthalpy recovery change of $270 \pm 10$ J g-atom$^{-1}$.

## 4. Conclusions

The peculiarity of the $Au_{49}Cu_{26.9}Si_{16.3}Ag_{5.5}Pd_{2.3}$ glass-forming system is that, during annealing below $T_g$, it exhibits at the microscopic scale a persistent and stationary regime of atomic dynamics, and at the macroscopic scale a hierarchical activation energy spectrum of enthalpy relaxation modes. Like many BMGs, $Au_{49}Cu_{26.9}Si_{16.3}Ag_{5.5}Pd_{2.3}$ is a multi-component system with large atomic size mismatches that result a dense atomic packing. However, in this system, the presence of a large amount of mobile atoms like Cu and Si contributes to the acceleration of secondary collective asymmetric diffusion processes to the point that intermediate relaxation processes branch off from the slow motion of the backbone structure represented by the sluggish diffusion of the large size atoms.

Our finding of a thermodynamic signature of a LLT below conventional $T_g$ implies that the frozen-in state during processing has retained the structure of the high-temperature fragile liquid, and that during annealing the system is driven to equilibrate into a new, stronger liquid. This process involves kinetics with relaxation times of the order of thousands of seconds and is found to be reversible. Our results are in agreement with observed LLT upon cooling of this system when the $T_g$ of the solidifying system is shifted to a lower temperature by applying an ultra-slow cooling rate [44]. The transition is reversible (i.e. strong-to-fragile) upon re-heating in DSC in the glass transition range, marked by a small value of the entropy change (~ 0.19 J g-atom$^{-1}$ K$^{-1}$). This value is as low as 2.4 % of the entropy of fusion of this system and it is the reason why the associated transition temperature is found for $Au_{49}Cu_{26.9}Si_{16.3}Ag_{5.5}Pd_{2.3}$ at a very low temperature, which is slightly below the glass transition temperature observed on the typical laboratory timescale.

This research was supported by the German Research Foundation (DFG) through Grant No. GA 1721/2-2. D. Cangialosi acknowledges the University of the Basque Country and Basque Country Government (Ref. No. IT-654-13 (GV)), Depto. Educación, Universidades e investigación; and Spanish Government (Grant No. MAT2015-63704-P, (MINECO/FEDER, UE)) for their financial support. We express our gratitude to C. Hafner Edemetall Technologie for the noble metals supply. We acknowledge and thank H. Vitoux and K.L. L'Hoste from the ESRF and W. Hembree for the support with the XPCS experiment, the CFM/MPC institute for the support with the flash calorimetry experiments, and G. Fiore for the experimental support with the melt spinning. We would like to thank especially R. Busch and O. Gross, S. Wei, M.H. Müser, S.V. Sukhomlinov, and W. Possart, for useful discussions.

**Materials and methods**

**Materials preparation.** To prepare $Au_{49}Ag_{5.5}Pd_{2.3}Cu_{26.9}Si_{16.3}$ glassy specimens, the mixture of elements (purity 99.995%) were melted and homogenized at a temperature of ~ 1100 K in an alumina crucible in an Indutherm MC15 casting apparatus and tilt-cast into their glassy state in a water-cooled Cu-mold. A few rods of 5 mm diameter and length of 34 mm and plates of 3x13x34 mm dimension were produced by applying identical tilt-casting procedures. Some of the rods were re-melted in a quartz tube and injected onto a rotating copper well under argon atmosphere conditions to obtain glassy ribbons of approximately 5 to 10 µm of thickness. Prior to all experiments, the specimens were shown to be x-ray amorphous by X-ray diffraction. To prevent room temperature aging, the material was stored in a freezer at roughly ~290 K.

**Calorimetry.** Conventional differential scanning calorimetry (DSC) was carried out under a constant argon flow in a power-compensated Perkin Elmer Hyper DSC 8500, equipped with an intracooler and calibrated according to the melting transitions of n-decane ($C_{10}H_{22}$), indium and tin. Approximately 200 mg of material was used for each DSC experiment. Fast scanning calorimetry (FSC) measurements were carried out in the Mettler Toledo Flash DSC 1. This was coupled with a temperature controller based on a two-stage intracooler. Calibration of the FSC was carried out according to the melting of indium at different rates. An onset value of $T_g$ = 396 K has been observed with a heating rate of $q_h$ = 0.333 K s$^{-1}$ for the bulk specimens as well for the ribbons. In cooling the $T_g$ was observed with the FSC at 439 K for $q_c$ = 5000 K s$^{-1}$, whereas for a $q_c$ = 0.333 K s$^{-1}$ it occurs at 402 K according to [45].

The enthalpy recovery data at 378 K and all those corresponding to $t_a \leq 100$ s for the other temperatures indicated in Fig. 1 are measured with the FSC using one sample for each annealing set. Each FSC specimen was directly placed onto the chip and the mass was estimated by comparing the heat of fusion obtained with this technique to that obtained with the conventional DSC with a known mass. A typical mass was in the range of 0.001 to 0.005 mg. Conventional DSC was used to acquire the enthalpy recovery data at longer $t_a$ and for the detection of the enthalpy released. In this latter case one sample was used for each experiment.

**Standard treatment.** Prior to the enthalpy relaxation experiments, a standard treatment was applied to sample by heating in with a rate of $q_h$ = 0.333 K s$^{-1}$ to a temperature of 418 K, which is above the end of the calorimetric glass transition and then cooled to 273 K with a rate of $q_c$ = 0.333 K s$^{-1}$. This assured the same enthalpic state for each specimen. After completion of annealing, the sample was first cooled to 273 K and then re-heated with the same rate to the end of the crystallization process for the detection of the enthalpy recovery. The crystalline baselines were produced by repeating the measurement in a second up-scan of the reacted material under identical conditions without removing the sample.

**Thermal step- protocol**. For this treatment the as-spun material was heated from RT by performing 3 h isothermal steps of 5 K up to 363 K and of 2 K up to 397 K, with a heating rate of 3 K/min. At the highest temperatures (between 383 and 397 K) the isotherms last ≈ 30-60 minutes. The specimen was subsequently cooled by performing isothermal steps of 0.5 K, with a cooling rate of 0.1 K/min and an annealing time of 1 to 4 h depending on the dynamics. This step- protocol was used to collect the XPCS data of Fig. 4, Fig. 6 and as pre-treatment for the DSC dashed-dotted up-scan curve of Fig. 1. In all cases the isotherms last a factor 5 to 40 longer than the corresponding relaxation time, thus assuring the correct evaluation of a stationary dynamics. For instance, for a τ(XPCS) ~200 s at 383 K, the isotherm lasts ≈ 4000 s, thus a factor 20 longer.

**Specific heat capacity measurement.** The $C_p(T)$ data of Fig. 2 were determined in DSC using a discontinuous step-method with $q_h$ = 0.333 K s$^{-1}$ and isothermal holding time of 120 s. The resulting step in the DSC heat flow of the sample was compared with that of a sapphire reference and empty measurement pan. The equations used to calculate the $C_p$ values from the heat flow signals are described elsewhere [37]. The specific heat capacities of the glassy and crystalline states were determined using aluminum pans in 10 K temperature intervals beginning from 193 K. The specific heat capacity of the equilibrium liquid above the melting point was determined using $Al_2O_3$ pans in 20 K intervals from 656 K up to 756 K. In a separate set of experiments, individual samples were heated in $Al_2O_3$ pans to a temperature of 723 K and undercooled in temperature intervals ranging from 20 K to 5 K. The sample was held at this temperature isothermally for 120 s and, in the case that the sample had not crystallized, the specific heat capacity of the supercooled liquid was determined using the method described in [75].

Additionally, the **$C_p$ on a long timescale** was estimated by determining the difference in specific heat capacities of the glassy state, $C_p(glass,T_i)$, and the equilibrium liquid, $C_p(liquid,T_i)$, at a temperature $T_i = T_1 + (T_2 - T_1)/2$, and approximating as $\Delta C_p(liquid-glass,T_i) \approx [\Delta H(tot,T_1) - \Delta H(tot,T_2)]/(T_2 - T_1)$, where $\Delta H(tot,T_1)$ and $\Delta H(tot,T_2)$ are the total endothermic enthalpy recovery, which was measured after equilibration at the specified temperature $T_1$ and $T_2$, respectively. This methodology was previously successfully applied [40,37]. The $C_p$ values at long timescale and the corresponding error bars are shown in Fig. 2.

**Viscosity relaxation measurement.** A thermomechanical analyzer (Netzsch TMA 402) was used to perform isothermal three-point beam-bending relaxation experiments below the glass transition. Amorphous beams with rectangular cross sectional areas of 0.3 to 1.1 mm² and a length of 13 mm were positioned on two sharp supporting edges. A load of 10 g was centrally applied by a silica probe with a wedge-shaped head and the samples were heated to the desired temperature with a constant rate of 0.333 K/s. There they were held isothermally at least until the end of the relaxation process while the beam deflection during relaxation was measured. The resulting viscosity (η) was calculated according to the Hagy equation

$$\eta = -\frac{g L^3}{144 \nu I_c}\left(M + \frac{\rho A L}{1.6}\right) \qquad [8],$$

where ν is the midpoint deflection rate (m/s) as measured in TMA, g is the gravitational constant (m/s²), L the support span of the apparatus (here $1.196 \times 10^{-2}$ m), $I_c$ the cross-section moment of inertia (m⁴), M the loading mass (kg), ρ the density of the sample (kg/m³) and A the cross-sectional area (m²).

**X-ray photon correlation spectroscopy.** The ribbon used for the XPCS was approximately 7 to 10 μm thick and was mounted in a resistively-heated Ni furnace with temperature stability better than 0.05 K. The experiment was carried out for a fixed incoming energy of 8.1 keV (λ = 1.53 Å), by using a partial coherent beam of 10x7 μm (HxV) with ≈ $10^{11}$ ph/s/200 mA. Speckles patterns were collected by a CCD detector (Andor Ikon-M, 13 μm pixel size) placed at ≈ 70 cm downstream of the sample, mounted horizontally in the scattering plane for an angle 2θ~39.6 degrees, thus measuring the dynamics in correspondence of the maximum of the structure factor $q_p$ = 2.78 Å$^{-1}$. The thermal step-protocol described above was applied to the sample while the time evolution of the microscopic dynamics is directly captured in XPCS by the intensity two-time correlation function $G(q,t_1,t_2)$ (TTCF). This quantity reflects the statistical similarity between speckles patterns measured at times $t_1$ and $t_2$ and is defined as

$$G(q, t_1, t_2) = \frac{\langle I(q,t_1)I(q,t_2)\rangle_p}{\langle I(q,t_1)\rangle_p \langle I(q,t_2)\rangle_p} \qquad [9],$$

where $<...>_p$ denotes an ensemble averaging performed on all pixels of the detector which in our case correspond to $q_p$. The time-averaged dynamics can be extracted from the TTCF over different time interval in order to get the standard intensity-intensity correlation function, $g_2(q,t) = <G(q,t_i,\Delta t)>$. This quantity can be directly related to the decay of density fluctuations in the glass, through the Siegert relation, $g_2(q,t) = 1 + \gamma|\Phi(q,t)|^2$, providing thus direct information on the dynamics. Here, γ is the experimental speckle contrast and Φ(q,t) = S(q,t)/S(q,0) is the normalized density-density correlation function. During the whole experiment, the experimental contrast was found ≈ 3% independently on the annealing time and temperature. Before and after each measurement at each temperature, we measured also the intensity profile in the [1.77-3.15] Å$^{-1}$ q-range in order to control possible structural evolution in the probed scale.

**Figures:**

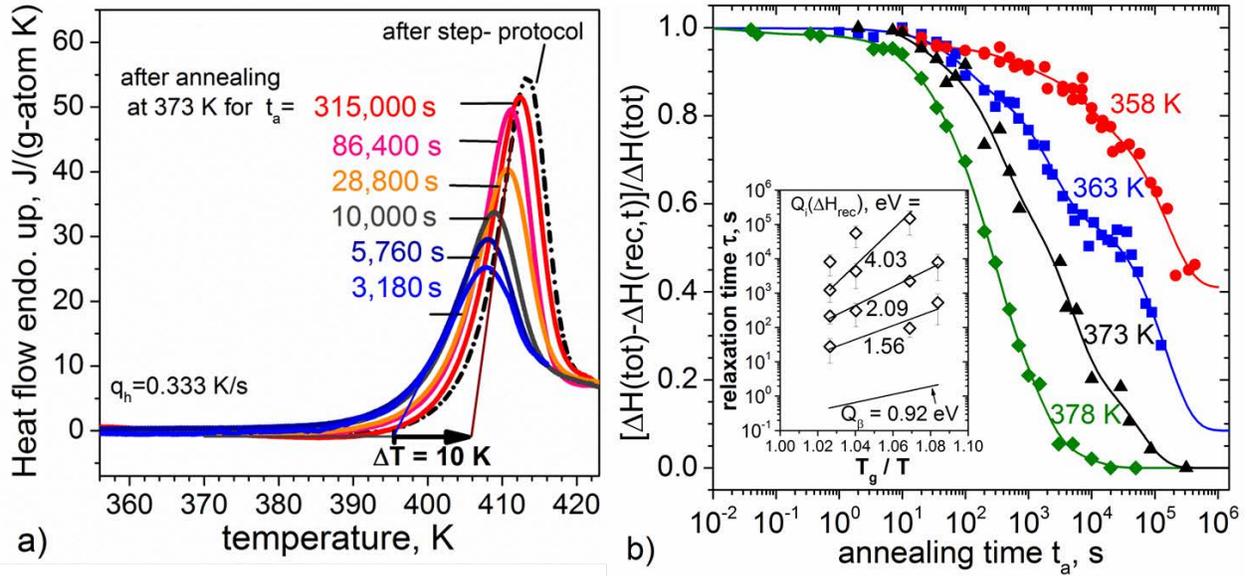

Fig. 1 Time evolution of the enthalpy change recovered back when the relaxed $Au_{49}Ag_{5.5}Pd_{2.3}Cu_{26.9}Si_{16.3}$ glass is heated up through the glass transition after annealing below $T_g$. a) Heat flow signals in the glass transition region of relaxed glasses after annealing at 373 K for the indicated annealing time $t_a$ (continuous curves) and of an as-spun specimen after a step-protocol (dashed-dot curve). The enthalpy change $\Delta H(rec,t)$ associated with the endothermic overshoot is obtained by integrating the area under signal with respect to the up-scan of a standard treated specimen. A thermodynamic crossover is identifiable after $t_a$ = 10,000 s as the onset of the endothermic overshoot signal shifts suddenly up to a maximum of 10 K, while the endothermic peak height accelerates in growth, with an increase of 70 J/g-atom. b) Plot of the normalized enthalpy recovery after annealing for the indicated temperatures as a function of $t_a$. $\Delta H(tot)$ is the maximum enthalpy change to reach equilibrium. Each data point corresponds to a single measurement sample. The continuous lines are the fits to Eq. 3. In the inset the $\tau_i(\Delta H_{rec})$ values of Eq. 3 are plotted versus $T_g$ (=396 K) scaled inverse temperature and are fitted to the Arrhenius equation with activation energy $Q_i$ and compared to the relaxation times from Ref. [45] for the β-relaxation.

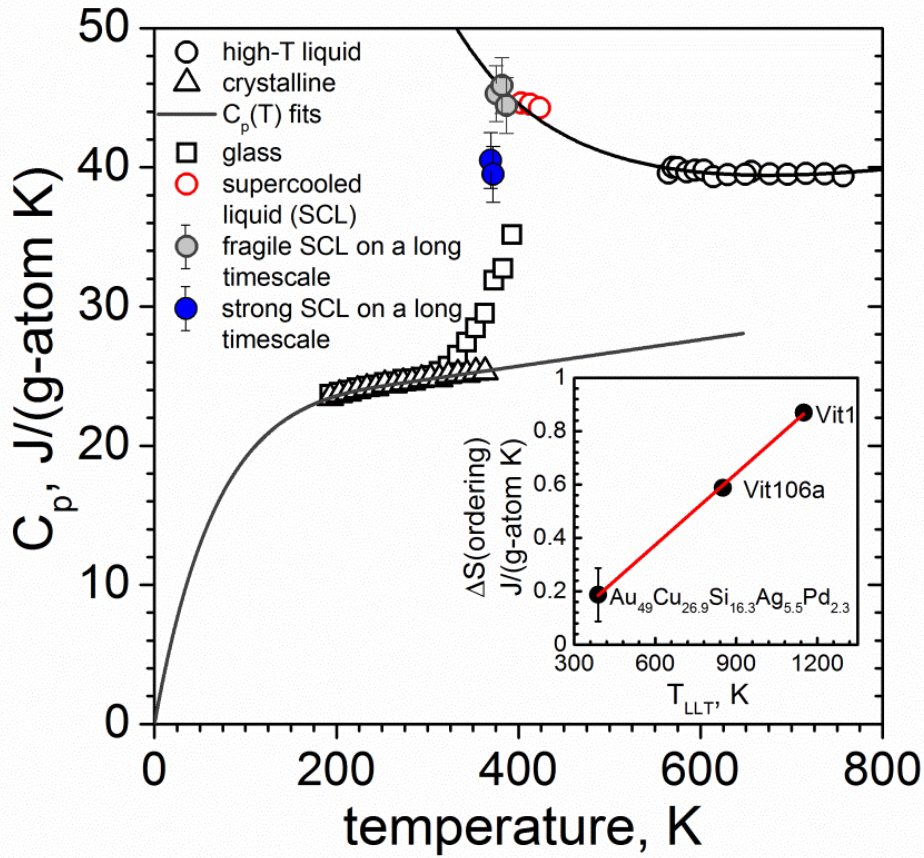

Fig. 2 Specific heat capacity $C_p(T)$ of $Au_{49}Ag_{5.5}Pd_{2.3}Cu_{26.9}Si_{16.3}$ determined in DSC using a step-method for the glass (squares), crystalline (triangles) and liquid (circles) states. The curves are the fits with Eqs. 4-6. The shaded circles with the error bars represent the values of $C_p$ for the supercooled liquid estimated from calorimetric experiments on a long timescale. For the other data, the error bars are smaller than the size of the symbols. Inset) Entropy change for the thermodynamic crossover as a function of the liquid-liquid transition temperature $T_{LLT}$ for the studied alloy in comparison with the Zr-based BMGs Vit 1 and Vit106a. The red line is a linear fit through the data.

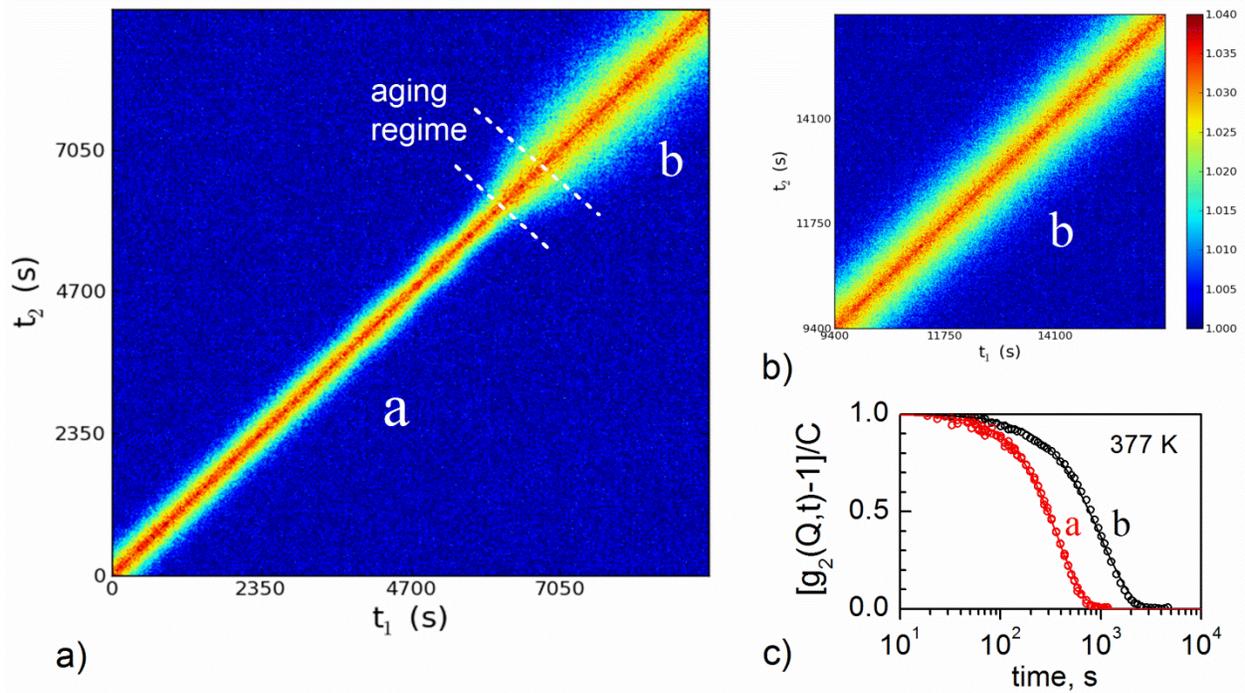

FIG. 3 Two-time XPCS correlation function (TTCF) for $Au_{49}Ag_{5.5}Pd_{2.3}Cu_{26.9}Si_{16.3}$ measured during isothermal annealing at 377 K for 5 h. a) TTCF during the first 9,400 s of annealing. b) TTCF for the last 7,000 s of annealing. The annealing followed a step-wise heating protocol from ambient temperature to 377 K, described in the methods. The axes correspond to the annealing time elapsed from the beginning of the data acquisition, while the width of the diagonal intensity profile is proportional to the microscopic structural relaxation time. c) Normalized one-time correlation functions extracted from the TTCF in the two stationary regimes labeled 'a' (from 0 to 6,000 s in Fig. 3a), and 'b' (for the entire correlation time in Fig. 3b). The solid lines through the data are fits to the Eq. 7, yielding $\tau_a = 570 \pm 26$ s and $\beta_a = 1.62 \pm 0.03$, and $\tau_b = 1650 \pm 7$ s and $\beta_b = 1.39 \pm 0.03$.

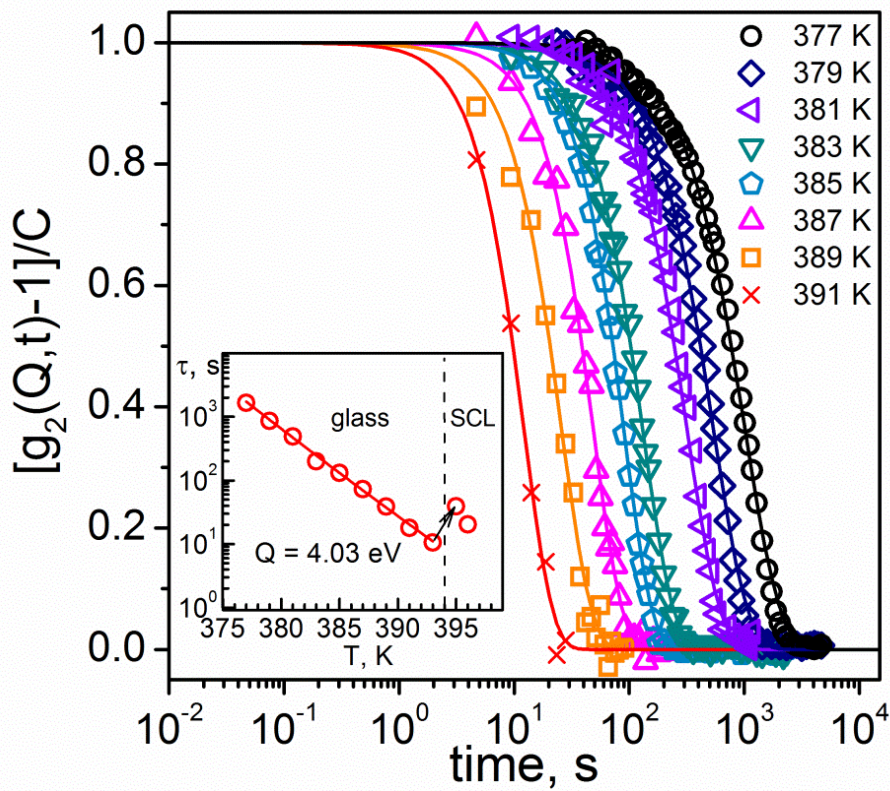

Fig. 4 Normalized intensity-intensity correlation functions for $Au_{49}Ag_{5.5}Pd_{2.3}Cu_{26.9}Si_{16.3}$ measured with XPCS at the indicated temperatures. a) The solid lines through the data are fits to the Eq. 7. The data at 377 K are for the stationary dynamics labeled 'b' of Fig. 3c. Inset) temperature evolution of the corresponding relaxation times. The continuous line is the Arrhenius-fit corresponding to an activation energy of Q = 4.03 ± 0.01 eV. The additional annealing at 395 and 396 K shows that the transition to the liquid state occurred during the heating step from 393 to 395 K.

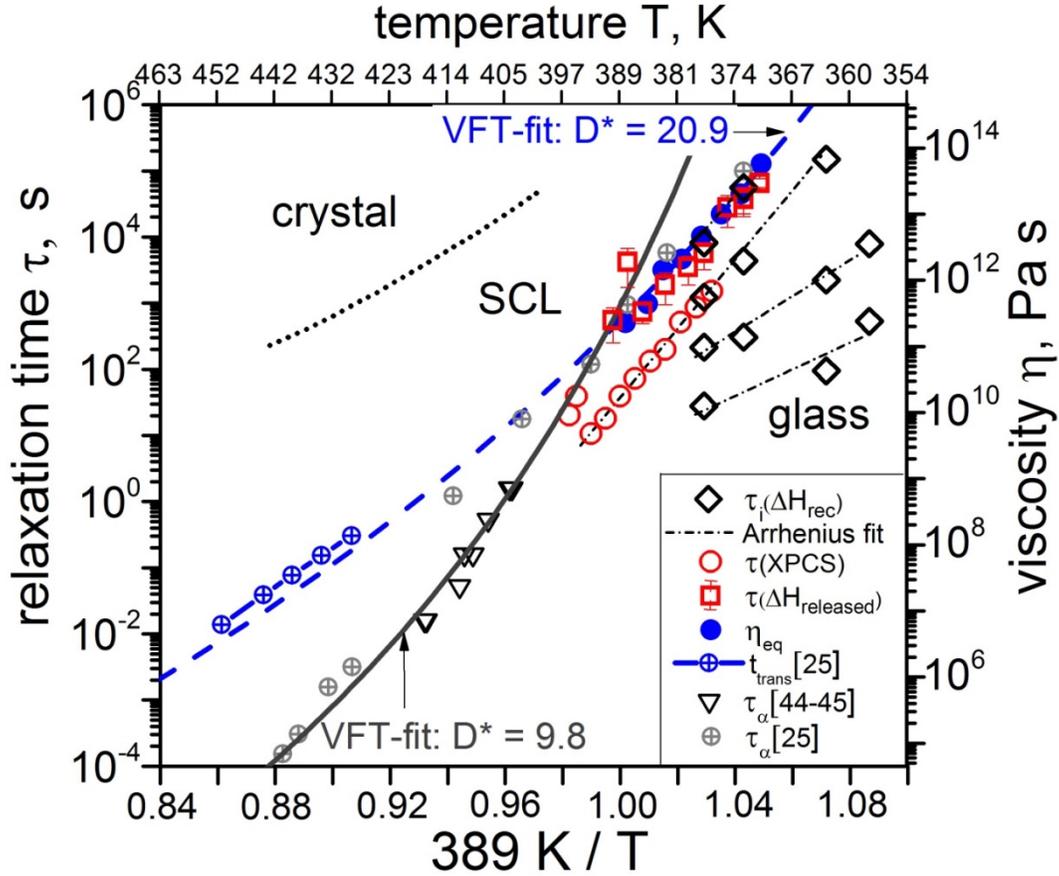

Fig. 5 Kinetic fragility diagram for $Au_{49}Ag_{5.5}Pd_{2.3}Cu_{26.9}Si_{16.3}$ of the relaxation times (open symbols) and equilibrium isothermal viscosities (filled blue circles, see SI) as a function of scaled inverse temperature. The scaling temperature is $T_{LLT}$ = 389 K from [44]. In detail, the various values of τ are represented by the diamonds for the $τ_i(ΔH_{rec})$ values of Fig. 1; the open circles for the for the XPCS data of Fig. 4; the squares for the equilibrium characteristic relaxation times $τ(ΔH_{released})$ (see SI); the open triangles for the frequency dependence of the DMA loss modulus from [44-45]; the crossed circles for the calorimetric data from [25]. The onset time of crystallization (dotted line) is in agreement with [26]. The dashed line is a fit of the viscosity data to the VFT equation (Eq. 1) with D* = 20.9 ± 0.4 and $T_0$ = 247 ± 3 K. The solid line is a VFT fit with D* = 9.8 ± 0.4 and $T_0$ = 311 ± 3 K. The dashed-dotted lines are Arrhenius fits with activation energy $Q_i$ from the enthalpy recovery spectrum analysis of Fig. 1.

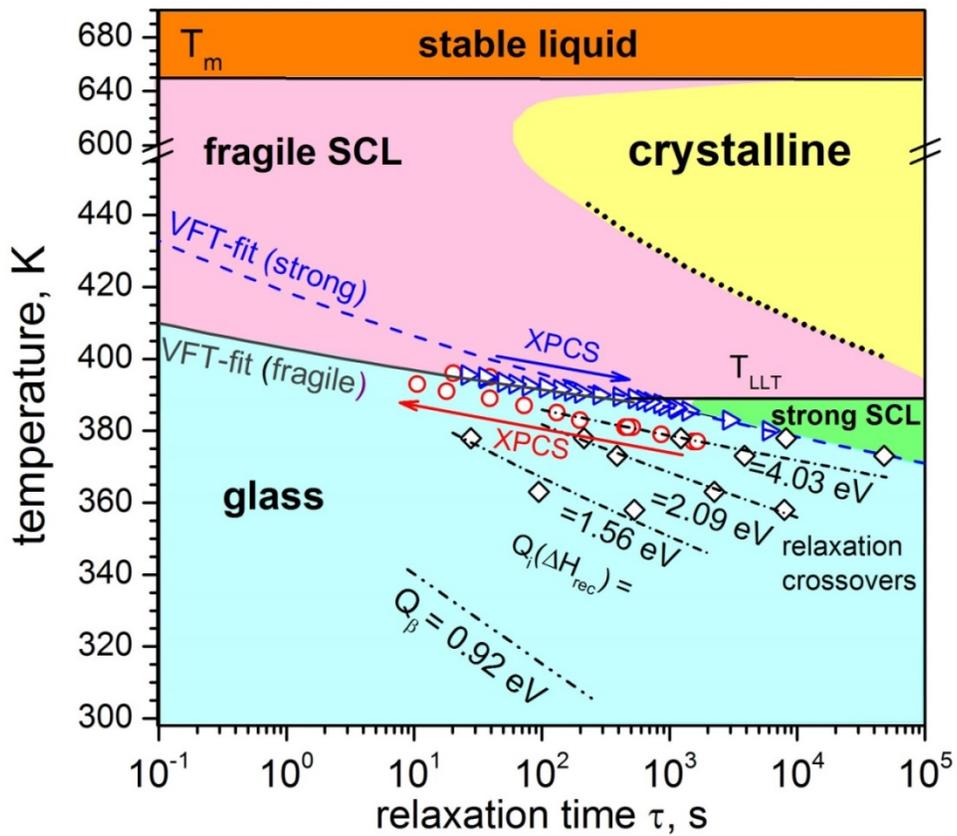

Fig. 6 Temperature-relaxation time-transformation diagram for $Au_{49}Ag_{5.5}Pd_{2.3}Cu_{26.9}Si_{16.3}$. The curves and the symbols correspond to the τ data as described in the legend of Fig. 5. The XPCS data in cooling (open blue triangles) are from [44].

**Supplementary information**

**1. Fitting parameters of the enthalpy recovery data analysis**

SI-Table 1 reports the fitting parameters for equation 3, used to model the enthalpy recovery data of Fig. 1b. The baseline A and amplitude values $h_i$ are unitless and correspond to values of [$\Delta H(tot)- \Delta H(rec,t)$]/$\Delta H(tot)$. The characteristic relaxation time values, $\tau_i$, are in units of seconds. The decays indicated with - where not possible to be identified because of the lack or the sparsity of data. The $Q_i$ values are the activation energy values of the individual enthalpy recovery processes obtained by fitting the $\tau_i(\Delta H(rec,T))$ values to the Arrhenius equation.

**2. Analysis of the isothermal enthalpy released**

In SI-Fig. 1 we report selected time evolution functions of the enthalpy release detected in DSC during relaxation of the standard treated $Au_{49}Ag_{5.5}Pd_{2.3}Cu_{26.9}Si_{16.3}$ glass. The calorimetric signals recorded during isothermal measurements are integrated as a function of annealing time. An example of heat flow for the annealing at 373 K is shown in the inset. Here, the relaxation process proceeds via a gradual exothermic heat release. The amount of enthalpy that is released from the system decreases with time during the relaxation until an equilibrium, but still metastable, thermodynamic liquid state is reached. The isothermal heat flow signal at 373 K show abnormal step-behavior after $t_w=1 \times 10^4$ s (see inset), which is the onset time of the final enthalpy recovery decay seen in Fig. 1. The overall signal was integrated as a function of time to obtain the $\Delta H(released,t)$ function, which is modeled with the single KWW equation as in [37]. The procedure was applied to various DSC isothermal signals performed between 368 and 390 K, and the resulting characteristic relaxation time values $\tau(\Delta H_{released})$ are plotted in Fig. 5 as open red squares.

**3. Isothermal viscosity change analysis**

In SI-Fig. 2 the time-evolution of the viscosity of the $Au_{49}Cu_{26.9}Si_{16.3}Ag_{5.5}Pd_{2.3}$ as measured by isothermal three-point beam-bending (ITPBB) is plotted. A thermomechanical analyzer (Netzsch TMA 402) was used to perform the ITPBB experiments and the detailed data analysis and fitting procedure with the KWW equation is described elsewhere, e.g. [29, 22]. It can be observed that, during the structural relaxation, the measured apparent viscosity rapidly rises and eventually starts to saturate towards an equilibrium value $\eta_{eq}$ at longer annealing times. The resulting $\eta_{eq}$ values are plotted as a function of annealing temperature in Fig. 5 as filled blue circles.

**4. Enthalpy diagram**

SI-Figure 3 is a diagram of the enthalpy change of the $Au_{49}Cu_{26.9}Si_{16.3}Ag_{5.5}Pd_{2.3}$ SCL as a function of temperature that is constructed to represent the enthalpy evolution of the glass during aging at 373 K. The reference state is the polycrystalline mixture; thus, $\Delta H(T) \equiv [H(liquid,T)-H(crystal,T)]$. For each experiment, in fact, the heat flow for the crystallized material is also measured and subtracted from the heat flow of the amorphous material. The black continuous curve corresponds to the enthalpy function of the fragile SCL determined by integrating the ΔC$_p$(T) of the high-T liquid, where ΔC$_p$(T)= C$_p$ (liquid,T)- C$_p$ (crystal,T) using Eqs. 4-5; namely,

$$\Delta H(T) = \Delta H_f - \int_T^{T_{liq}} \Delta C_p(T')dT, \qquad [SI-2]$$

where the ΔH$_f$ and T$_{liq}$ are the experimentally detected heat of fusion and liquidus temperature, respectively, reported in SI-Table 2.

Prior to the DSC aging experiments, the as-cast material is pre-treated following a standard procedure that cools from the SCL at 418 K with a constant rate q$_c$ = 0.333 K s$^{-1}$ (see methods) This ensures the same enthalpic state for every glassy sample. The samples were then reheated with 0.333 K s$^{-1}$ to the isothermal holding temperature. During this standard treatment, a certain amount of relaxation (enthalpy release) occurred and the effective initial enthalpic state is indicated in SI- Fig. 3 with the label 'standard treated glass'. The experimentally determined amount of enthalpy recovered after the second-to-last decay equals the expected enthalpy change between the initial state of the glass (red dashed line) and the extrapolated enthalpy of the high-T liquid (black dotted line), whereas the total amount of enthalpy that is recovered back after the final equilibration is somewhat greater. We attribute this excess enthalpy, to the exothermic latent heat of the fragile-to-strong transition in the ultra-viscous state. For the annealing at 373 K, this latent heat is approximately 70 ± 25 J g-atom$^{-1}$ K$^{-1}$, which corresponds to an entropy change of 0.187 J g-atom$^{-1}$.

At the end of the isothermal measurement, the sample was cooled to 273 K before it was reheated for the detection of the enthalpy recovery overshoot signal. As a consequence of this cooling procedure, we freeze-in the final relaxed enthalpic state, which differs depending on the applied annealing temperature and time. In SI-Fig. 3 this is represented for the annealing at 373 K by the enthalpy value of the blue dash-dotted line. The annealing temperature can thus be considered as the fictive temperature of the frozen-in system after a long annealing. The temperature-dependence of the enthalpy of the strong SCL (blue continuous line) is obtained by adjusting the slope of the strong SCL curve to the slope of the fragile SCL and to the ΔC$_p$-ratio of the two liquids using the following relation:

$$\frac{dH(strong\ SCL)}{dT} = \frac{\Delta C_p(strong\ SCL)}{\Delta C_p(fragile\ SCL)} * \frac{dH(fragile\ SCL)}{dT} \qquad [SI-3]$$

Eq. SI-3 is valid under the assumption that, at least within a small enough temperature interval, the temperature dependence of the enthalpy change can be described with a linear equation.

SI-Table 1: Results of the enthalpy recovery spectrum analysis for the $Au_{49}Ag_{5.5}Pd_{2.3}Cu_{26.9}Si_{16.3}$. a) Fitting parameters for the equation: $f(t) = A + \sum_{i=1}^{N} h_i \exp(-t/\tau_i)$ [Eq. 3] used to fit the data of Fig. 1 for each annealing temperature. b) The activation energy values $Q_i$ of the individual enthalpy recovery decay. $Q_1$ and $Q_5$ could not be determined based on only two values of τ.

| a) | at 358 K | at 363 K | at 373 K | at 378 K |
|---|---|---|---|---|
| $\tau_1$, s | 9.86 | - | - | 0.041 |
| $\tau_2$, s | 533.16 | 95.23 | - | 27.82 |
| $\tau_3$, s | 7921.12 | 2247.39 | 306.37 | 215.61 |
| $\tau_4$, s | 157741.00 | 149293.00 | 4369.41 | 1226.95 |
| $\tau_5$, s | - | - | 55972.78 | 8165.86 |
| $h_1$ | 0.093 | - | - | 0.015 |
| $h_2$ | 0.042 | 0.120 | - | 0.117 |
| $h_3$ | 0.110 | 0.295 | 0.295 | 0.433 |
| $h_4$ | 0.397 | 0.498 | 0.417 | 0.371 |
| $h_5$ | - | - | 0.243 | 0.065 |
| A | 0.41 | 0.085 | 0 | 0 |
| ΔH(tot), kJ/g-atom | 0.65 | 0.55 | 0.27 | 0.18 |

| b) | $Q_2$, eV | $Q_3$, eV | $Q_4$, eV |
|---|---|---|---|
| | 1.56 | 2.09 | 4.03 |

SI-Table 2: Thermodynamic parameters for the $Au_{49}Ag_{5.5}Pd_{2.3}Cu_{26.9}Si_{16.3}$.

| a) Calorimetric values measured with a rate of 0.333 K s$^{-1}$ | | | | | | |
|---|---|---|---|---|---|---|
| $T_{g,onset}$ K | $T_x$ K | $T_{eut}$ K | $T_{liq}$ K | $\Delta H_x$ kJ/g-at | $\Delta H_f$ kJ/g-at | $\Delta S_f = \Delta H_f/T_{liq}$ J/g-at K |
| 396 | 456 | 615 | 647 | -5.59 | 5.14 | 7.94 |

| b) $C_p$-fitting parameters: a, b, c, d, and M. $T_D$ is the Debye-T and $T_K$ is the Kauzmann-T | | | | | | |
|---|---|---|---|---|---|---|
| a×10$^{-3}$ J/g-at·K$^2$ | b×10$^6$ J·K/g-at | c×10$^{-3}$ J/g-at·K$^2$ | d×10$^{-6}$ J/g-at·K$^3$ | M | $T_D$ K | $T_K$ K |
| 13.410 | 2.4410 | -14.0036 | 43.3036 | 1.0005 | 102.93 | 318 |

SI-Figures:

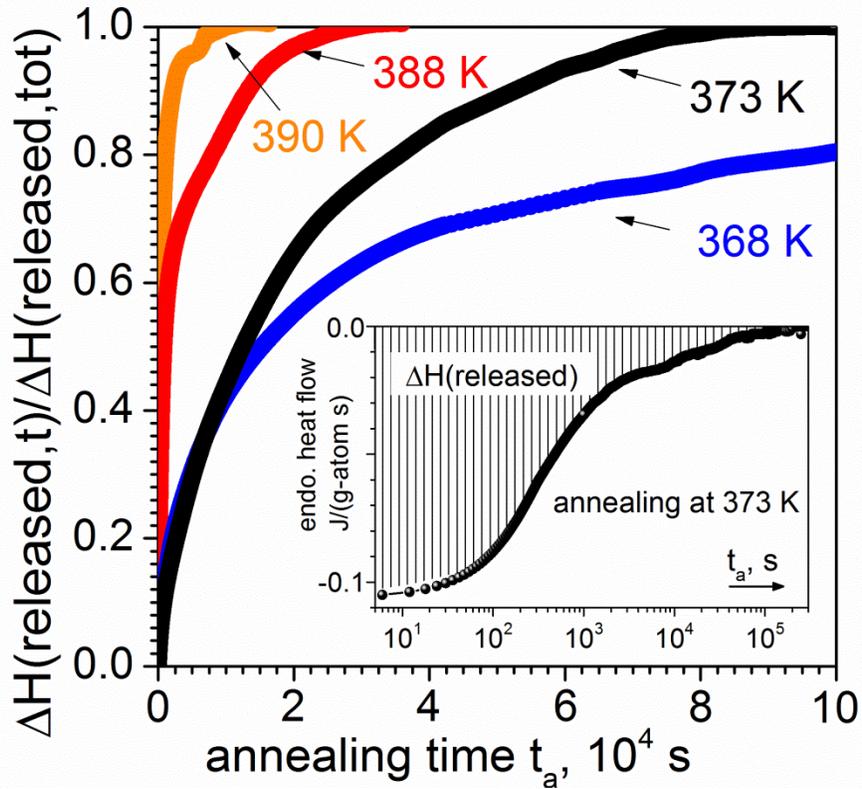

SI-Fig. 1 Normalized enthalpy released curves $\Delta H(t)/\Delta H_{total}$ for the standard treated $Au_{49}Ag_{5.5}Pd_{2.3}Cu_{26.9}Si_{16.3}$ BMG for the indicated isothermal annealing. With decreasing annealing temperature a longer time for equilibration is observed. Inset) Heat flow change during the annealing at 373 K subtracted by the heat flow change of a fully crystallized sample. The shaded area represents the total enthalpy that was released during aging.

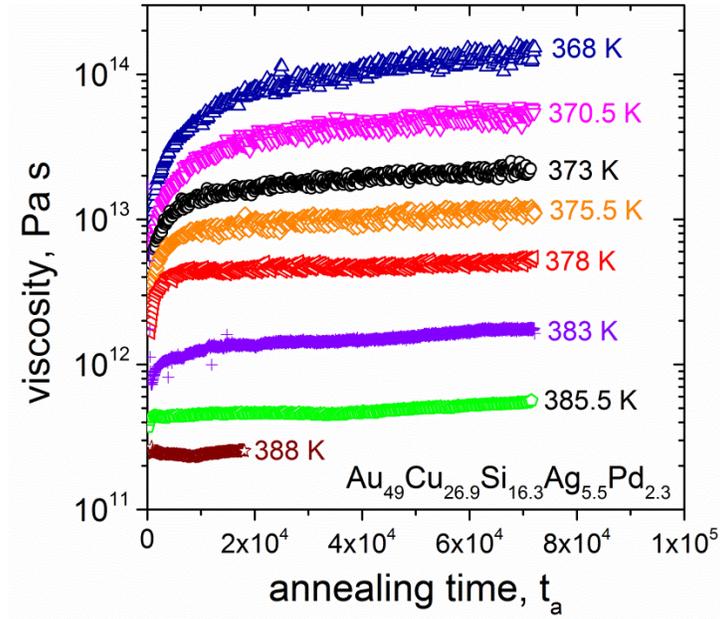

SI-Fig. 2 Viscosity change measured using isothermal three points beam bending for the as-cast Au$_{49}$Ag$_{5.5}$Pd$_{2.3}$Cu$_{26.9}$Si$_{16.3}$ BMG.

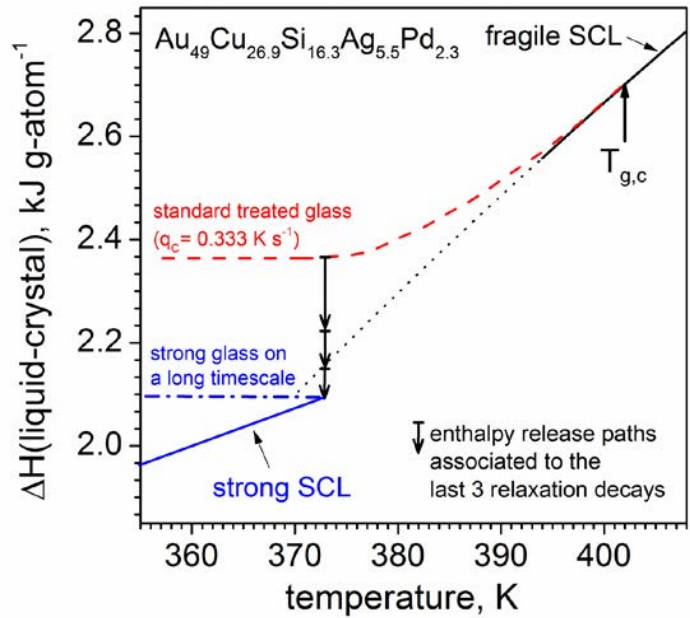

SI-Fig. 3 Enthalpy diagram in the vicinity of the glass transition for the Au$_{49}$Cu$_{26.9}$Si$_{16.3}$Ag$_{5.5}$Pd$_{2.3}$ supercooled liquid in reference to the crystalline mixture. The T$_{g,c}$ = 402 K is the glass transition temperature during the cooling with q$_c$ = 0.333 K s$^{-1}$ [45]. The downward arrows represent the multistep path behavior of the enthalpy equilibration.